\documentclass{elsart} 
\usepackage{epsfig}
\begin{document}

\emph{Physica A (2005) to appear.}

\begin{frontmatter}
\title{Melting Transition of Directly-Linked Gold 
Nanoparticle DNA Assembly}

\author{Y.\ Sun, N.\ C.\ Harris, and C.-H.\ Kiang$^*$}
\address{Department of Physics and Astronomy\\
Rice University, Houston, TX\ \ 77005--1892}

\begin{abstract}
DNA melting and hybridization is a fundamental biological process
as well as a crucial step in many modern biotechnology applications.
DNA confined on surfaces exhibits different behavior from that
in free solutions.  The system of DNA-capped gold nanoparticles 
exhibits unique phase transitions and represents a new class
of complex fluids.  Depending on the sequence of the DNA, 
particles can be linked to each other through direct complementary
DNA sequences or via a ``linker'' DNA whose sequence is complementary
to the sequence attached to the gold nanoparticles.
We observed different melting transitions for these two distinct systems.
\end{abstract}

\begin{keyword}
DNA phase transition \sep gold nanoparticle  \sep DNA melting

\PACS 82.39.Pj \sep 87.15.He \sep 87.68.+z \sep 87.15.-v
\end{keyword}

\end{frontmatter}


\section{Introduction}

Melting of the DNA duplex is the process by which two DNA strands unbind
upon heating.  The nature of this transition has been studied for
decades \cite{Wartell85a,Hwa97a,Breslauer99a}.  For short DNA with fewer 
than 12--14 base pairs, melting and hybridization can be described by
a two-state model as an equilibrium between single- and
double-stranded DNA \cite{Crothers00a,CantorII}.  For long and
heterogeneous DNA, the melting curve exhibits a multi-step behavior
consisting of plateaus with different sizes separated by sharp jumps.
Although much of the thermodynamic properties of the melting of free
DNA are known, DNA melting in a constrained space, such
as on surfaces, is still poorly understood \cite {Magnasco02a}. DNA
molecules functionalized with gold nanoparticles provide a model
system for such study.  

The sequence-specific hybridization properties of DNA
have been used for self-assembly of nanostructures 
and for highly sensitive DNA detection \cite{Mirkin96a,Kiang03a}.  
Previous work relies on
a linker DNA \cite{Mirkin96a,Kiang03a,Kiang05a,Kiang05b}, and it
has been suggested that entropic cooperativity plays an important
role in the sharp phase transition of such DNA-linked nanoparticle
assembly systems.  On the other hand, most simulations do not 
explicitly incorporate linker DNA \cite{Stroud03a}, and the results
cannot be directly compared to experimental data.  Here we 
synthesized a system that eliminated the usage of a linker DNA
and found that the melting transitions of these direct-linked 
gold particles exhibit distinct behavior from those connected
via a linker DNA.

\section{Experimental Procedures}

\begin{figure}[h]
\begin{center}
\epsfig{file=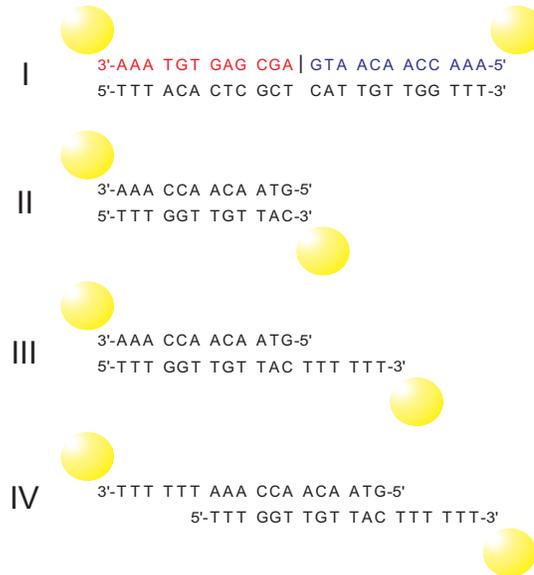,height=3.0in,clip=}
\end{center}
\caption{DNA sequences used to form DNA-linked gold nanoparticles.
Sample I is connected through a linker DNA.  The line between bases
A and G in the probe DNA sequences indicates that there is no
chemical bond between these two bases.  Sample II-IV are directly
connected through surface-attached DNA with spacings of 12, 18, and
24 DNA bases between particles.}
\label{fig:fig1}
\end{figure}

Sample was prepared according to the procedures described 
in \cite{Kiang03a}.  Briefly, 
DNA-capped gold nanoparticles were prepared by conjugating gold
colloidal nanoparticles with thiol-modified DNA. 
The configuration of the DNA used in different experiments is
illustrated in Fig.~\ref{fig:fig1}.  
We prepared four sets of samples with different DNA lengths and sequences.
In sample I, the gold particles are connected through a 24-base DNA linker; 
in sample II, the gold particles are directly connected via 12-base 
complementary DNA on gold particles; in sample III, the gold
particles are directly connected via 12- and 18-base DNA; in sample IV, 
the gold particles are directly connected via 18-base DNA.  

The aggregates of DNA-linked gold colloids were allowed to stand at 4
$^\circ$C for several days for aggregation.  Optical spectroscopy was
used to study the phase transition of the DNA-linked gold colloids,
since DNA bases have strong absorption in the UV region
\cite{Crothers00a,CantorII}.  We monitor the thermal melting by
measuring the extinction at 260 nm while slowly heating the solution
containing aggregates.  The solution was heated 
from 25 to 75 $^\circ$C at a rate of 0.5 $^\circ$C/min.
All spectra were taken with a PerkinElmer Lambda 45 spectrophotometer
equipped with a peltier temperature controller, magnetic stirrer, and
a temperature probe.  The recorded temperature of the sample was 
measured by a temperature probe.

\section{Results and Discussion}

\begin{figure}[tb]
\begin{center}
\epsfig{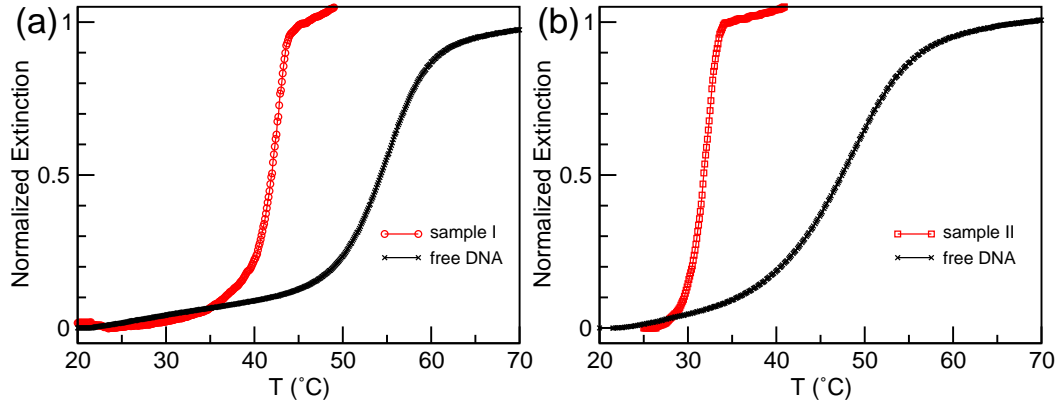}
\end{center}
\caption{Melting curves of gold nanoparticles connected (a) with a DNA
linker (sample I), and (b) via direct hybridization of complementary
surface-attached DNA (sample II).  The corresponding free DNA melting
curves are also shown.}
\label{fig:fig2}
\end{figure}

Fig.~\ref{fig:fig2} shows the the melting curves of sample I (a) 
and sample II (b).  The melting curves of corresponding DNA in solution
are also shown.  The melting temperature of DNA duplex attached 
to gold particle surfaces is lower than that of free DNA, and the melting
transition is much sharper.   However, direct comparison of melting temperatures
between these two systems is difficult, since the DNA compositions
are different for these two systems.  The system without linker appears to have
a lower melting temperature, perhaps due to the short distance between
gold particles (12 base versus 24 base).  

\begin{figure}[tb]
\begin{center}
\epsfig{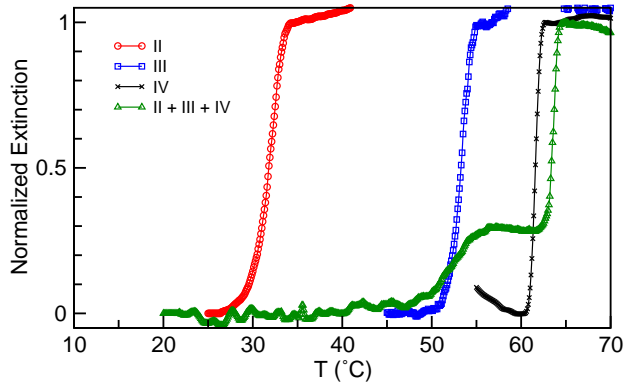}
\end{center}
\caption{Melting curves of 12/12 (sample~II), 12/18 (sample~III), 
18/18 (sample~IV), and a combination of 12/12, 12/18, and 
18/18 (Sample II~+~III~+~IV).  The mixed system shows multi-step melting
at temperatures corresponding to the $T_m$'s, within the experimental uncertainty,
for Sample~II and Sample~III.}
\label{fig:fig3}
\end{figure}

To study how the melting depends on the spacing between gold particles,
we prepared gold particles capped with 18 base DNA (Sample III), 
which is composed of identical sequence to the 12 base DNA in 
Sample II plus a 6 base DNA spacer (see Fig.~\ref{fig:fig1}).
The difference between sample II and III is the spacer DNA length, 
which alter the spacing between gold particles.  
The melting temperature is 62 $^\circ$C for sample IV 
versus 32 $^\circ$C for sample II (see Fig~\ref{fig:fig3}),
which suggests that increased particle spacing leads to higher
melting temperature of the assemblies.

To introduce disorder, we mixed the 12 and 18 base DNA capped gold
particles in one solution.  Thus, the 12 base DNA are allowed to
hybridize with either 12 or 18 base DNA, and the 18 base to 18 or 12
base DNA.  This combination allows three possible duplex formations:
12/12 (sample I), 12/18 (sample II), and 18/18 (sample III)
hybridization in one solution and possibly in one aggregate.  Note
that in all three base pairing only 12 bases are complementary, and
the only variable is the non-pairing DNA spacer length, which controls
the inter-particle distance.  Since the duplexes with higher melting
temperatures are more stable, we expect to see more of those duplexes
forming.  Indeed, Fig.~\ref{fig:fig3} shows that the most abundant 
duplex is the 18/18 combination, followed by 12/18, with 
almost no 12/12 duplex formed.

\begin{figure}[tb]
\begin{center}
\epsfig{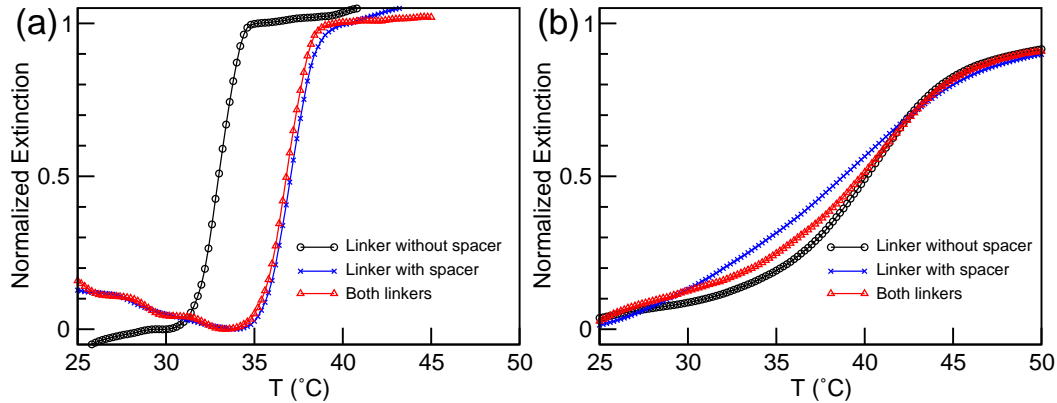}
\end{center}
\caption{Melting curves of DNA duplex containing spacer for (a) nanoparticle
assembly, and (b) free DNA.}
\label{fig:fig4}
\end{figure}

The multi-step melting is an unusual phenomenon in DNA-capped gold particle
assembly.  For the system connected by either 24 or 30 base DNA
linker, where the 30 base linker differs from the 24 base by an extra 6
base spacer in the middle of the linker, 
heating the assembly results in a single melting temperature $T_m$. 
The $T_m$ of the system with spacer is higher (37~$^\circ$C) than that 
without spacer (33~$^\circ$C).  When equal amounts of linkers with 
and without spacer are present in the solution, the system has a $T_m$
in between the high $T_m$ (37~$^\circ$C) and low $T_m$ (33~$^\circ$C) systems.  
The $T_m$ of the mixed system (36.5~$^\circ$C) is much closer to the 
more stable system (37~$^\circ$C), as illustrated
in Fig.~\ref{fig:fig4}a.  However, free DNA
with the same sequences exhibits different trend in $T_m$
(see Fig.~\ref{fig:fig4}b), where a linker with spacer results in 
lower melting temperature than that without spacer.
The finding suggests that the
inter-particle distance plays an important role in determining the $T_m
$ in the nanoparticle system.  
On the other hand, the multi-step melting in
the system without linker DNA suggests most clusters are composed of either 
12/18 connections or 18/18 connectionis and few with both connected in the 
same cluster, unlike the systems with linker DNA.
The abundance of clusters of a given connection is
related to its stability.  We speculate that once the cluster nucleate 
with a certain type of connection (defined here by the DNA length, hence the
interparticle spacing) only the same type of
connection is allowed to grow.  Further studies are needed to determine
whether this phenomenon is kinetics or thermodynamics driven.

\section{Summary}

In summary, we have studied the thermal denaturation
of DNA strands attached to gold nanoparticle surfaces.  In the DNA-capped
gold nanoparticle systems, the interactions are complex, involving
DNA-DNA interactions and particle-particle interactions.  The DNA are
constrained to a gold particle surface and often exhibit interesting
behaviors not seen by DNA in free solution.  The multi-step melting 
of systems with different spacers is unique to the systems directly 
linked with DNA that are attached to gold nanoparticles.

$^*$To whom correspondence should be addressed, email: chkiang@rice.edu.

\bibliography{nolinker}

\end{document}